\documentclass[superscriptaddress,showpacs,preprintnumbers,amsmath,amssymb,twocolumn,floats]{revtex4}
\usepackage{txfonts}
\usepackage{amssymb}
\usepackage[pdftex]{graphicx}
\usepackage{sidecap}
\usepackage{CJK}
\usepackage{txfonts}

\begin{document}

\title{Drastic electronic structure reconstruction of Ca$_{1-x}$Pr$_x$Fe$_2$As$_2$ in the collapsed tetragonal phase}

\author{D. F. Xu}
\affiliation{State Key Laboratory of Surface Physics, Department of Physics, and Laboratory of Advanced Materials, Fudan University, Shanghai 200433, People's Republic of China}
\author{D. W. Shen}
\email{dwshen@mail.sim.ac.cn}
\affiliation{State Key Laboratory of Functional Materials for Informatics, Shanghai Institute of Microsystem and Information Technology (SIMIT), Chinese Academy of Sciences, Shanghai 200050, People's Republic of China}
\author{J. Jiang}
\author{Z. R. Ye}
\author{X. Liu}
\author{X. H. Niu}
\author{H. C. Xu}
\author{Y. J. Yan}
\author{T. Zhang}
\author{B. P. Xie}
\author{D. L. Feng}
\email{dlfeng@fudan.edu.cn}
\affiliation{State Key Laboratory of Surface Physics, Department of Physics, and Laboratory of Advanced Materials, Fudan University, Shanghai 200433, People's Republic of China}

\begin{abstract}
We report the electronic structure reconstruction of Ca$_{1-x}$Pr$_x$Fe$_2$As$_2$ ($x$ = 0.1 and 0.15) in the low temperature collapsed tetragonal (CT) phase observed by angle-resolved photoemission spectroscopy. Different from Ca(Fe$_{1-x}$Rh$_x$)$_2$As$_2$ and the annealed CaFe$_2$As$_2$ where all hole Fermi surfaces are absent in their CT phases, the cylindrical hole Fermi surface can still be observed in the CT phase of Ca$_{1-x}$Pr$_x$Fe$_2$As$_2$. Furthermore, we found at least three well separated electron-like bands around the zone corner in the CT phase of Ca$_{1-x}$Pr$_x$Fe$_2$As$_2$, which are more dispersive than the electron-like bands in the high temperature tetragonal phase. Based on these observations, we propose that the weakening of correlations (as indicated by the reduced effective mass), rather than the lack of Fermi surface nesting, might be responsible for the absence of magnetic ordering and superconductivity in the CT phase.
\end{abstract}


\maketitle

\section{INTRODUCTION}

The interplay of magnetism, lattice and superconductivity is an important issue in iron-based superconductors. Typically, the ground state of iron pnictides is a magnetically ordered collinear-antiferromagnetic (CAF) state {\cite{pnictide_SDW1, pnictide_SDW2}}. Carrier doping or the application of hydrostatic pressure can gradually suppress the CAF order and then lead to the emergence of a superconducting dome in the phase diagram {\cite{induce_SC1, induce_SC2, induce_SC3}}. On the other hand, under 0.35~GPa hydrostatic pressure, CaFe$_2$As$_2$ was found to undergo a remarkable structural transition to a collapsed tetragonal (CT) phase at low temperatures {\cite{CaFe2As2_hydro1, CaFe2As2_hydro2, CaFe2As2_hydro3}}. Intriguingly, this CT phase, characterized by the shrinkage in the $c$-direction by approximately 10\% without breaking the symmetry, is non-magnetic and non-superconducting. As shown by calculations, neutron scattering, and x-ray emission spectroscopy, Fe local moments are absent in the CT phase {\cite{non-magnetic_theory1, non-magnetic_theory2, non-magnetic_experiment1, non-magnetic_experiment2, CaFe2As2_hydro3}}. Therefore, it provides a unique platform for studying a non-superconducting iron-pnictide system without Fe local moments. From a different angle of view, this would facilitate the understanding of the electronic structure which is relevant to superconductivity and local moments in the superconducting iron pnictides.

Angle-resolved photoemission spectroscopy (ARPES) is one of the most direct methods of studying the electronic structure of solids. However, reports on the CT phase had been limited by the necessity of external pressure. Early on, since the non-magnetic phosphide CaFe$_2$P$_2$ is a close structural analog of CaFe$_2$As$_2$ in the CT phase, its electronic structure is considered to somewhat reflect the electronic behavior of the CT phase {\cite{CaFe2P2_QO}}. The de Haas-van Alphen study showed that the Fermi surface of CaFe$_2$P$_2$ is highly three-dimensional and the hole pocket only exists around the $Z$ point of the Brillouin zone, which are very different from the electronic strucure of an iron pnictide superconductor. Recently, the CT phase was found to be stabilized under ambient pressure through introducing chemical pressure by specific dopants or thermal treatment {\cite{CaFe2As2_doping, CaFe2As2_thermal_treatment}}. Subsequent ARPES experiments on Ca(Fe$_{1-x}$Rh$_x$)$_2$As$_2$ {\cite{Ca(FeRh)2As2_CT_ARPES}} and the annealed CaFe$_2$As$_2$ {\cite{CaFe2As2_CT_ARPES_1, CaFe2As2_CT_ARPES_2}} have revealed significant differences in the electronic structure between the high temperature tetragonal (HT) phase and the CT phase. The predicted vanishment of the hole pocket at the zone center {\cite{non-magnetic_theory1, CaFe2As2_no_hole_1, CaFe2As2_no_hole_2}} has been confirmed in both cases. The Fermi surface nesting between electron and hole Fermi surfaces thus does not exist anymore, which was argued to be responsible for the absence of magnetic fluctuation and superconductivity in the CT phase {\cite{Ca(FeRh)2As2_CT_ARPES, CaFe2As2_CT_ARPES_1}}.

Ca$_{1-x}$Pr$_x$Fe$_2$As$_2$ is one of the several iron pnictides that undergo the collapsed tetragonal transition under ambient pressure upon cooling {\cite{CaFe2As2_hydro3, CaFe2As2_doping, CaPr_2SC_phase, CaPr_anisotropy, CaPr_NMR, CaPr_Hoffman, CaPr_local_inhomogeneity}}. However, little is known about its electronic structure to date. To examine the common properties of the band reconstruction across the CT transition, it is worthwhile to study the low-lying electronic structure of Ca$_{1-x}$Pr$_x$Fe$_2$As$_2$, which might provide clues to understand the interplay of magnetism, lattice and superconductivity in iron-based superconductors.

In this article, we report a detailed ARPES study on Ca$_{1-x}$Pr$_x$Fe$_2$As$_2$ single crystals ($x$ = 0.1 and 0.15). The electronic structure in the high temperature tetragonal (HT) phase resembles that of the parent compound CaFe$_2$As$_2$ in its paramagnetic state. Across the CT transition, bands around the zone center shift towards different directions. However, the cylindrical hole Fermi surface can still be observed in its CT phase. This is different from the CT phases of Ca(Fe$_{1-x}$Rh$_x$)$_2$As$_2$ and the annealed CaFe$_2$As$_2$, in which all hole Fermi surfaces are absent. At least three well separated electron-like bands can be resolved around the zone corner in the CT phase. This contradiction indicates that the absence of magnetic ordering and superconductivity in the CT phase is not correlated with the presence or absence of hole pocket. Our polarization dependence data show strong orbital mixing for the bands in the CT phase. Illustrated by the detailed temperature dependence data, band reconstruction occurs abruptly across the transition, and the hysteresis of reconstruction in the temperature cycle further confirms its first order nature. By comparing the effective masses of the bands in these two phases, we propose that the suppression of electronic correlation, rather than the lack of Fermi surface nesting, might be responsible for the absence of magnetic fluctuation and superconductivity in the CT phase { \cite{ZRY}}. Our results indicate that the electronic structure of the CT phase of Ca$_{1-x}$Pr$_x$Fe$_2$As$_2$ are rather different from those of other known CT compounds, and thus might help facilitate a more comprehensive understanding of the CT phase.

\section{SAMPLE PROPERTIES AND EXPERIMENTAL SETUP}

\begin{figure}[t!]
\includegraphics[width=8.6cm]{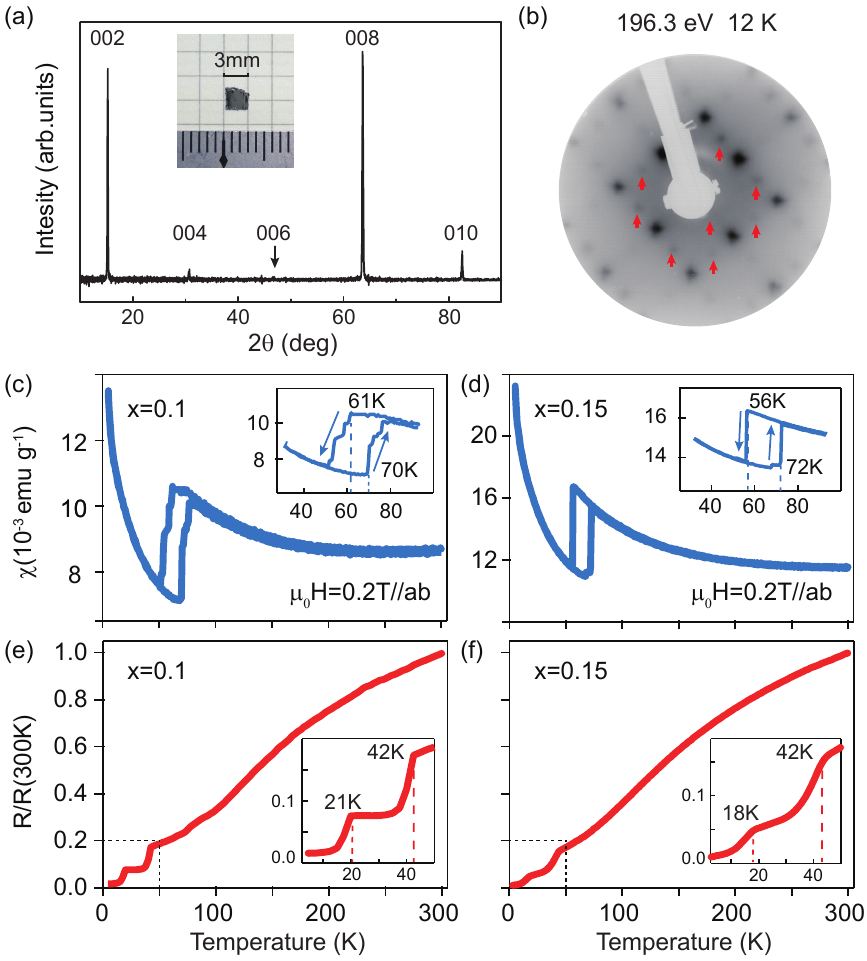}
\caption{(Color online) The crystallographic and transport properties of Ca$_{1-x}$Pr$_x$Fe$_2$As$_2$. (a) The X-ray diffraction pattern of a $x$ = 0.1 single crystal. The inset is the photography of a typical sample. (b) The low-energy electron diffraction pattern of the $x$ = 0.1 sample in the CT phase at 15~K, showing 1$\times$2 and 2$\times$1 surface reconstructions marked by red arrows. (c), (d) DC susceptibility as a function of temperature for the $x$ = 0.1 and 0.15 samples under a magnetic field of 0.2~T respectively. Data around the hysteretic drops are magnified in the insets, which correspond to the CT transition. (e), (f)  The temperature dependence of the electrical resistivity for the $x$ = 0.1 and 0.15 samples respectively. Data at low temperatures are magnified in the insets.}
\label{lattice}
\end{figure}

High quality Ca$_{1-x}$Pr$_x$Fe$_2$As$_2$ single crystals were synthesized by the FeAs self-flux method as described elsewhere {\cite{CaFe2As2_growth}}. The typical dimension of our samples is 3$\times$2.5$\times$0.1~mm$^3$ as shown by the inset of Fig.~\ref{lattice}(a). The typical X-ray diffraction (XRD) pattern of our sample [Fig.~\ref{lattice}(a)] agrees well with the previous report {\cite{CaPr_2SC_phase}}. We have studied samples with two compositions ($x$ = 0.1 and 0.15), and the electron probe micro-analysis (EPMA) gives actual chemical compositions of Ca:Pr:Fe:As = 0.967:0.152:2.000:2.010 for $x$ = 0.1, and 0.931:0.231:2.000:2.024 for $x$ = 0.15 samples (normalized to the stoichiometric value of Fe), respectively. These samples show slight differences in the onset CT transition temperatures, as illustrated by the hysteretic drops in magnetic susceptibility [Figs.~\ref{lattice}(c) and (d)]. Moreover, there are two drops [Figs.~\ref{lattice}(e) and (f)] in resistivity near 20~K and 40~K for both samples, indicating the possible existence of two superconducting phases {\cite{CaPr_2SC_phase, CaPr_local_inhomogeneity}}. However, no evident diamagnetic behavior could be identified in the magnetic susceptibility in the zero-field cool mode, which suggests the filamentary nature of the superconductivity in this material {\cite{CaPr_local_inhomogeneity}}. These properties are qualitatively consistent with previous reports on Ca$_{1-x}$Pr$_x$Fe$_2$As$_2$ {\cite{CaFe2As2_doping, CaPr_local_inhomogeneity, CaPr_NMR}}. There is no evident difference in electronic structure between $x$ = 0.1 and $x$ = 0.15 samples except for slight changes in the chemical potential ({\textless} 10~meV), thus only representative data will be shown for them.

ARPES measurements were performed at (i) Beamline 5-4 of Stanford Synchrotron Radiation Lightsource (SSRL), (ii) the SIS beamline of the Swiss Light Source (SLS) and (iii) an in-house system equipped with a SPECS UVLS helium discharge lamp. VG-Scienta R4000 electron analyzers are equipped in all setups. The angular resolution was 0.3$^\circ$ and the overall energy resolution was better than 15~meV. Samples were cleaved $in$-$situ$ and measured under ultra-high vacuum better than 5$\times$10$^{-11}$ torr. The cleaved sample surface exhibits 1$\times$2 and 2$\times$1 reconstructions [Fig.~\ref{lattice}(b)], which were frequently observed in cleaved $A$Fe$_2$As$_2$ compounds {\cite{122_surface1, 122_surface2, CaPr_Hoffman}}. Recent scanning tunneling microscopy study on Ca$_{1-x}$Pr$_{x}$Fe$_2$As$_2$ identify the 1$\times$2 and 2$\times$1 surface as a half-Ca termination {\cite{CaPr_Hoffman}}, which preserves the charge neutrality. This is a must for the surface to be representative of the bulk.

\section{EXPERIMENTAL RESULTS}

\subsection{Band structures in the HT phase and the CT phase}

\begin{figure}[t!]
\includegraphics[width=8.6cm]{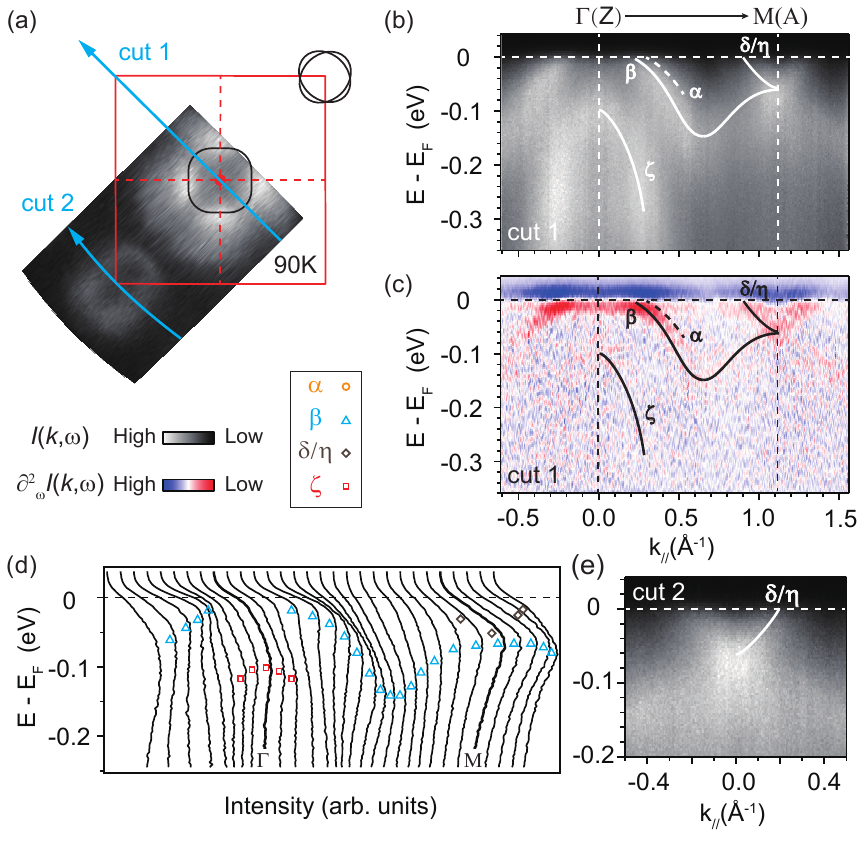}
\caption{(Color online) Electronic structure of Ca$_{1-x}$Pr$_{x}$Fe$_2$As$_2$ ($x$ = 0.1) in the HT phase. (a) The photoemission intensity map integrated over [$E_F$ - 5~meV, $E_F$ + 5~meV]. Solid black lines represent the Fermi surface contours. Cut directions are indicated by blue arrows. (b), (c) The photoemission intensity plot and its second derivative with respect to energy along cut 1. (d) The selected EDCs for data in panel (b). (e) The photoemission intensity plot along cut 2. Solid lines in panel (b), (c) and (e) are guide of eye for band dispersions. Data were taken with randomly-polarized 21.2~eV photons at 90~K.}
\label{HT}
\end{figure}

The electronic structure of Ca$_{1-x}$Pr$_{x}$Fe$_2$As$_2$ ($x$ = 0.1) in the HT phase is presented in Fig.~\ref{HT}. The Fermi surface consists of a large square hole pocket around the zone center and two concentric elliptical electron pockets at the zone corner as illustrated in Fig.~\ref{HT}(a). The photoemission intensity plot and its second derivative with respect to energy along cut 1 [indicated in Fig.~\ref{HT}(a)] are shown in Figs.~\ref{HT}(b) and (c), respectively. At the zone center, two bands (assigned as ${\beta}$ and ${\zeta}$) can be identified by tracking the peaks in the corresponding energy distribution curves (EDCs) in Fig.~\ref{HT}(d). The ${\beta}$ band crosses the Fermi level ($E_F$) while the band top of ${\zeta}$ is located at about -100~meV. In this photon energy, the hole-like band ${\alpha}$ is hard to resolve directly due to the overwhelming intensity of the broad ${\beta}$ band. At the zone corner along cut 2 [Fig.~\ref{HT}(e)], one broad electron-like feature can be resolved. This feature should be contributed by two electron-like bands (assigned as ${\delta}$ and ${\eta}$), which could be better distinguished in other ${k_z}$ planes and would be shown later. Overall, the band structure of Ca$_{1-x}$Pr$_{x}$Fe$_2$As$_2$ ($x$ = 0.1) in the HT phase resembles that of Ca$_2$Fe$_2$As$_2$ in the paramagnetic state reported by previous ARPES experiments {\cite{Chenfei_ARPES, CaFe2As2_ARPES}}.

\begin{figure}[t!]
\includegraphics[width=8.6cm]{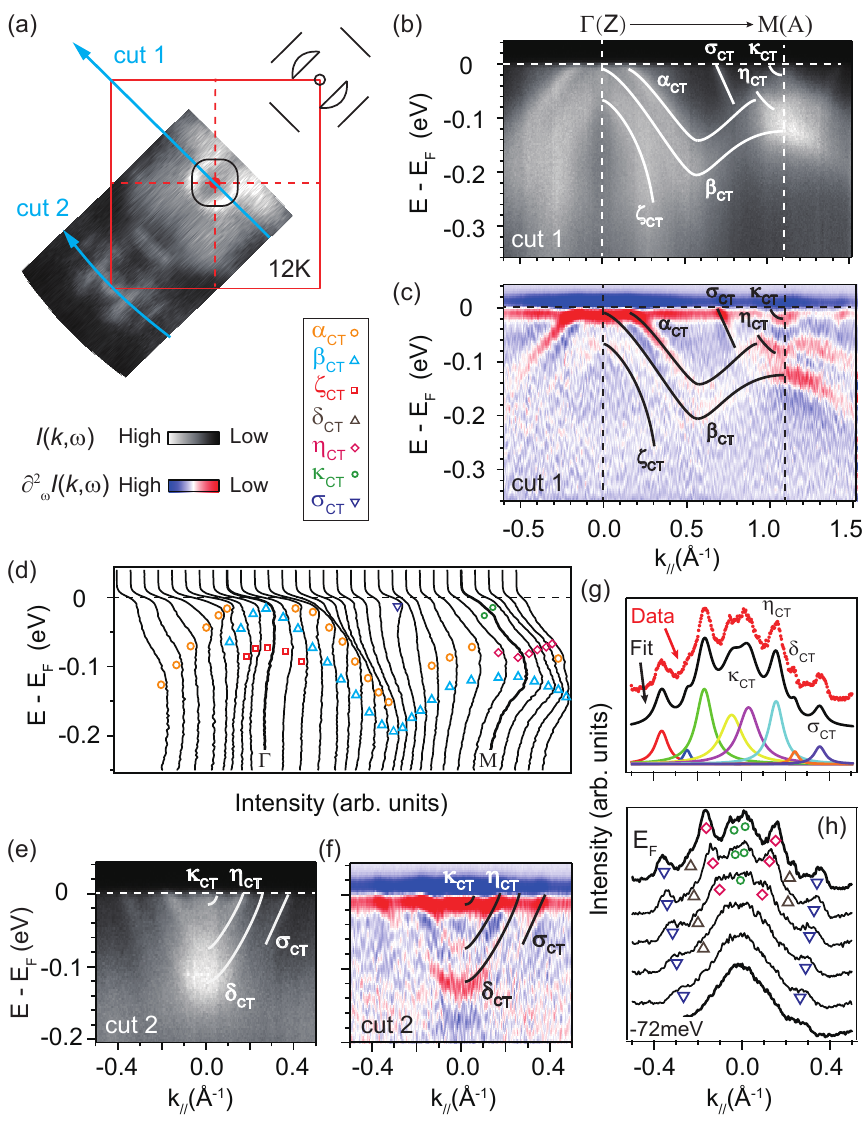}
\caption{(Color online) Electronic structure of Ca$_{1-x}$Pr$_{x}$Fe$_2$As$_2$ ($x$ = 0.1) in the CT phase. (a) The photoemission intensity map integrated over [$E_F$ - 5~meV, $E_F$ + 5~meV]. Solid black lines represent the Fermi surface contours. Cut directions are indicated by blue arrows. (b), (c) The photoemission intensity plot and its second derivative with respect to energy along cut 1. (d) The selected EDCs for data in panel (b). (e), (f) The photoemission intensity plot and its second derivative with respect to energy along cut 2. (g) The MDC integrated over [$E_F$ - 6~meV, $E_F$ + 6~meV] fitted by 8 Lorentzian peaks. (h) The corresponding MDCs near $E_F$ in panel (e). Solid lines in panel (b), (c), (e) and (f) are guide of eye for band dispersions. Data were taken with randomly-polarized 21.2~eV photons at 12~K.}
\label{LT}
\end{figure}

\begin{figure}[t!]
\includegraphics[width=8.6cm]{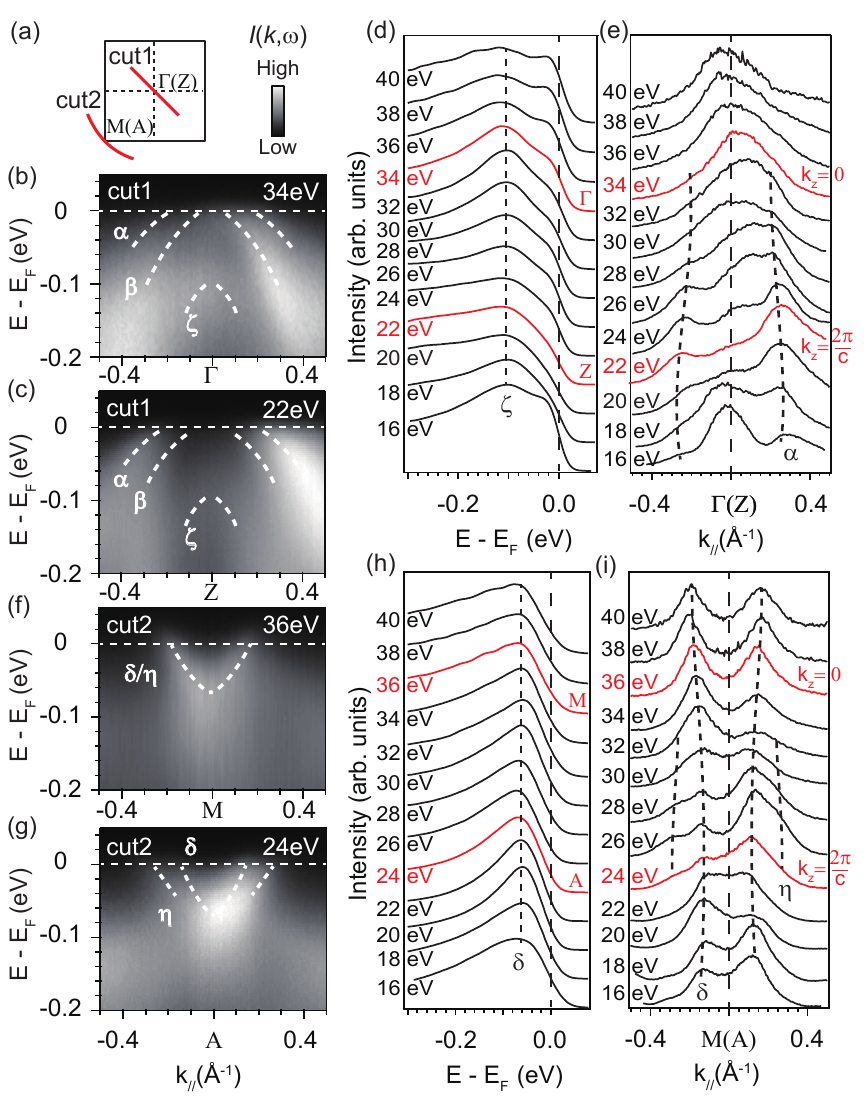}
\caption{(Color online) Photon-energy dependence of bands of Ca$_{1-x}$Pr$_{x}$Fe$_2$As$_2$ ($x$ = 0.15) in the HT phase at 115~K. (a) Indication of cut directions in the projected two-dimensional Brillouin zone. (b), (c) The photoemission intensity plots along cut 1 taken with 34~eV and 22~eV photons respectively. (d) The photon-energy dependence of the EDCs at ${\varGamma}$($Z$). (e) The photon-energy dependence of the MDCs at $E_F$ along cut 1. (f), (g) The photoemission intensity plots along cut 2 taken with 36~eV and 24~eV photons respectively. (h) The photon-energy dependence of the EDCs at $M$($A$). (e) The photon-energy dependence of the MDCs at $E_F$ along cut 2. Dashed lines in panel (b), (c), (f), (g) are guide of eye for band dispersions.}
\label{KZH}
\end{figure}


The Fermi surface topology and band structure of Ca$_{1-x}$Pr$_{x}$Fe$_2$As$_2$ ($x$ = 0.1) change dramatically in the CT phase as shown in Fig.~\ref{LT}. The reconstructed Fermi surface consists of a smaller square hole pocket around the zone center and several rather complicated electron pockets at the zone corner [Fig.~\ref{LT}(a)]. The disconnection of the Fermi surface contours might be due to the matrix element effects. The photoemission intensity plot and its second derivative with respect to energy along cut 1 are shown in Figs.~\ref{LT}(b) and (c), respectively. Three bands around the zone center (assigned as ${\alpha_{CT}}$, ${\beta_{CT}}$ and ${\zeta_{CT}}$) could be resolved. From the EDCs shown in Fig.~\ref{LT}(d), ${\alpha_{CT}}$ crosses Fermi level while the band top of ${\beta_{CT}}$ is located at -10~meV. The fast dispersing ${\zeta_{CT}}$ may originate from ${\zeta}$, and its band top moves from -100~meV to about -70~meV.

Around the zone corner, the photoemission intensity plot and its second derivative with respect to energy along cut 2 are plotted in Figs.~\ref{LT}(e) and (f), respectively. Eight Fermi crossings can be clearly distinguished by the fitting of the momentum distribution curves (MDCs) near $E_F$ as indicated in Fig.~\ref{LT}(g), six of which belong to three electron-like bands (assigned as ${\delta_{CT}}$, ${\eta_{CT}}$ and ${\kappa_{CT}}$), which can be further confirmed by tracking the peaks in MDCs below $E_F$ in Fig.~\ref{LT}(h). However, it is hard to determine whether the band with the largest crossings (assigned as ${\sigma_{CT}}$) is a hole-like band around the zone center or an electron-like band around the zone corner. Moreover, faint intensity is detected around the $X$ point (not shown here). This additional intensity is consistent with the folding of bands which would be induced from the 1$\times$2 and 2$\times$1 reconstructions at the surface {\cite{122_surface2}}. The main feature of the electronic structure of the CT phase of Ca$_{1-x}$Pr$_{x}$Fe$_2$As$_2$ is that bands tend to be more dispersive in general. For example, around the zone corner, the Fermi velocity of ${\delta_{CT}}$, ${\eta_{CT}}$ and ${\sigma_{CT}}$ is larger than those of ${\delta}$ (${\eta}$) in the HT phase, indicating that the electronic correlations are suppressed in the CT phase.

\begin{figure}
\includegraphics[width=8.6cm]{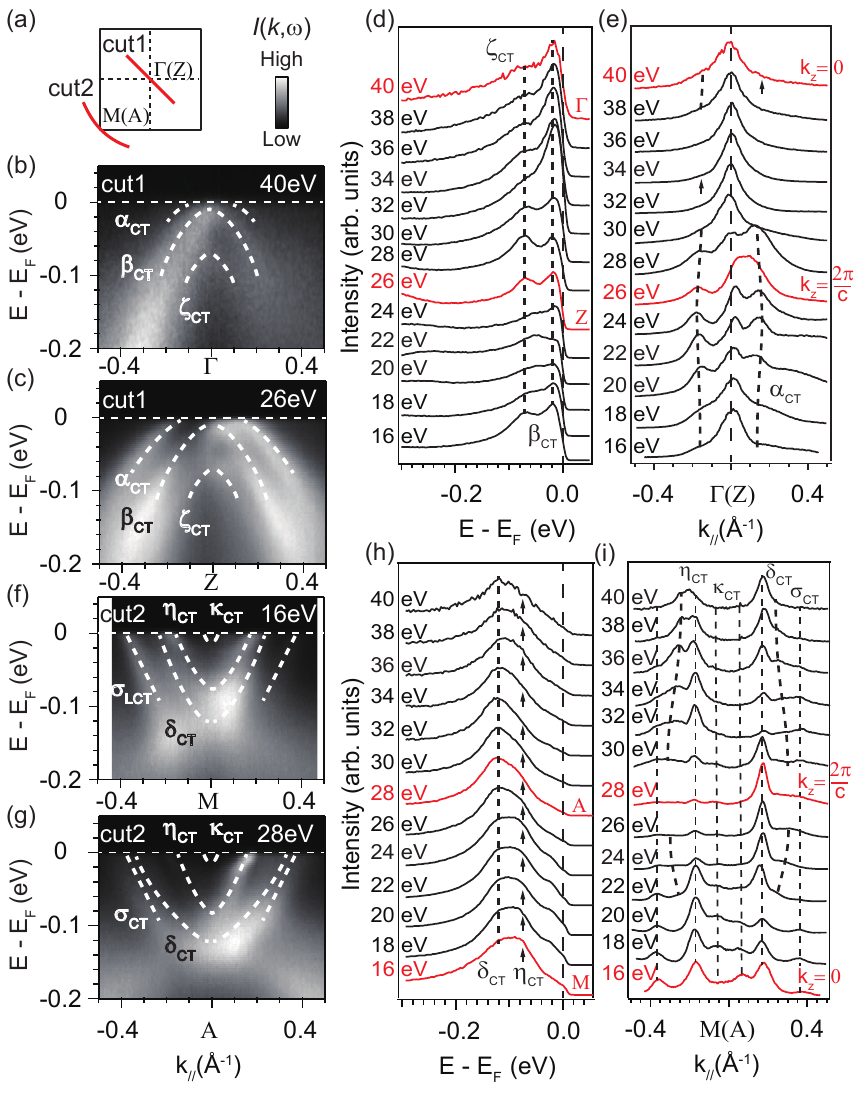}
\caption{(Color online) Photon-energy dependence of bands of Ca$_{1-x}$Pr$_{x}$Fe$_2$As$_2$ ($x$ = 0.15) in the CT phase at 6~K. (a) Indication of cut directions in the projected two-dimensional Brillouin zone. (b), (c) The photoemission intensity plots along cut 1 taken with 40~eV and 26~eV photons respectively. (d) The photon-energy dependence of the EDCs at ${\varGamma}$($Z$). (e) The photon-energy dependence of the MDCs at $E_F$ along cut 1. (f), (g) The photoemission intensity plots along cut 2 taken with 16~eV and 28~eV photons respectively. (h) The photon-energy dependence of the EDCs at $M$($A$). (e) The photon-energy dependence of the MDCs at $E_F$ along cut 2. Dashed lines in in panel (b), (c), (f), (g) are guide of eye for band dispersions.}
\label{KZL}
\end{figure}


All the above data were taken with 21.2~eV photons. However, the sizable contraction of lattice parameter $c$ across the transition can result in the expansion of the reciprocal lattice along $k_z$ direction, which means that the data taken in the HT and CT phase lie in different $k_z$ planes. If the electronic structure of Ca$_{1-x}$Pr$_{x}$Fe$_2$As$_2$ is highly three-dimensional, it is probable that the differences observed above are merely related to $k_z$ dispersions. To rule out this possibility, photon energy dependent ARPES experiments were performed. Our data cover more than half of the Brillouin zone along $k_z$ direction. Considering the periodicity of the band dispersions along $k_z$ direction, we use an inner potential of 16~eV to determine the corresponding high symmetry points.

Photon energy dependence of the bands in the HT phase are presented in Fig.~\ref{KZH}. At the zone center, two representative photoemission intensity plots taken with different photon energies are compared in Figs.~\ref{KZH}(b) and (c). The intensities of ${\alpha}$ and ${\beta}$ show anticorrelation with each other. While ${\beta}$ is at its strongest around ${\varGamma}$ point [Fig.~\ref{KZH}(b)], ${\alpha}$ is mostly enhanced around $Z$ point [Fig.~\ref{KZH}(c)]. However, there is rather little variation in the band dispersions. For instance, the band tops of ${\zeta}$ show negligible $k_z$ dependence as illustrated in Fig.~\ref{KZH}(d). As for ${\alpha}$, the MDCs at $E_F$ in Fig.~\ref{KZH}(e) show the slight changes in the Fermi crossings upon variation in photon energy, indicative of its moderate $k_z$ dependence. The intensity of ${\alpha}$ become faint around the ${\varGamma}$ point, which is probably due to the differences in photoemission matrix element. Around the zone corner [Figs.~\ref{KZH}(f)-(i)], the broad electron-like band splits into two bands at specific photon energies as illustrated in Figs.~\ref{KZH}(g) and (i). The broad peaks in EDCs around -70~meV correspond to the band bottom of ${\delta}$, and their positions do not change with photon energies.

\begin{figure*}[t!]
\includegraphics[width=17.8cm]{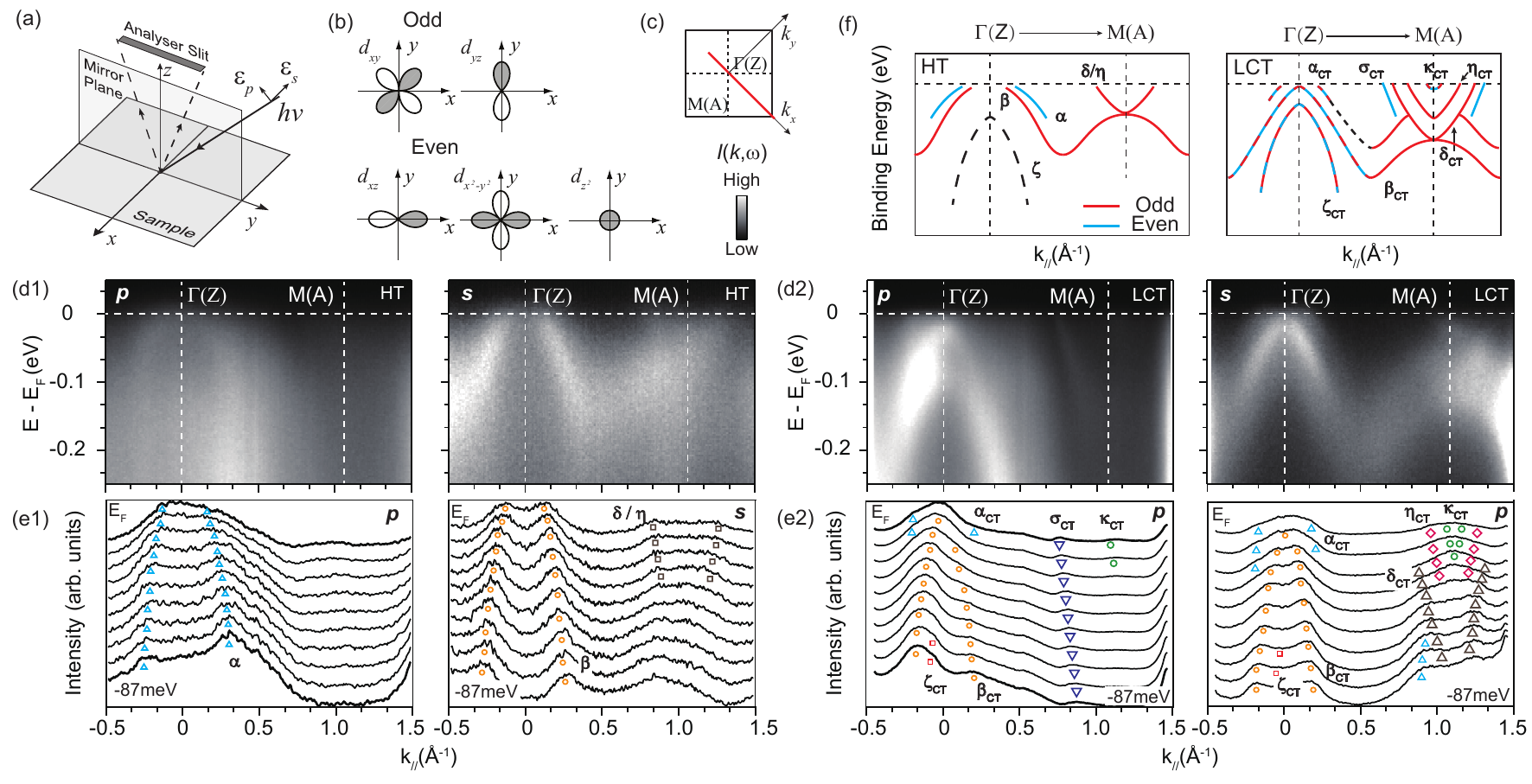}
\caption{(Color online) Polarization dependence of photoemission data of Ca$_{1-x}$Pr$_{x}$Fe$_2$As$_2$ ($x$ = 0.15). (a) Experimental setup for polarization-dependent ARPES. (b) Illustration of the spatial symmetry of the 3$d$ orbitals with respect to the mirror plane. (c) Indication of cut direction in the projected two-dimensional Brillouin zone. (d1) The photoemission intensity plots taken with 103~eV linearly-polarized photons along $\varGamma$($Z$)-$M$($A$) direction in the HT phase at 90~K. The left one was taken under the $p$ geometry while the right one under the $s$ geometry. (e1) The corresponding MDCs for the data in panel (d1). (d2), (e2) The corresponding CT phase (T = 16~K) data for data in panel (d1) and (e1) respectively. Data were taken with 72~eV linearly-polarized photons. (f) The summary of the orbitals (even or odd) of low-energy electronic structure in these two phases. Bands that could not be clearly resolved are represented by dashed black lines.}
\label{polar}
\end{figure*}

In the CT phase (Fig.~\ref{KZL}), the representative photoemission intensity plots taken with different photon energies are shown in Figs.~\ref{KZL}(b), (c), (f) and (g). As illustrated in Figs.~\ref{KZL}(b)-(e), the intensities of ${\alpha_{CT}}$, ${\beta_{CT}}$ and ${\zeta_{CT}}$ behave almost the same as those of ${\alpha}$, ${\beta}$ and ${\zeta}$ in the HT phase, except their band tops are located at different binding energies, as mentioned above. At the zone corner [Figs.~\ref{KZL}(f)-(i)], the band bottoms of ${\delta_{CT}}$ and ${\eta_{CT}}$ do not vary with photon energies [Fig.~\ref{KZL}(h)]. In Fig.~\ref{KZL}(i), the peak positions of ${\kappa_{CT}}$, ${\eta_{CT}}$ and ${\sigma_{CT}}$ in MDCs at $E_F$ show negligible $k_z$ dependence. While the Fermi crossing of ${\delta_{CT}}$ varies slightly, indicating its moderate $k_z$ dispersion.

The observed photon energy dependency indicates that the electronic structure of Ca$_{1-x}$Pr$_x$Fe$_2$As$_2$ is quite two-dimensional in general. As a result, it rules out the possibility that the differences between the electronic structures of the HT and CT phase are caused by the complication related to $k_z$ dispersions. Moreover, the band tops of ${\zeta}$, ${\zeta_{CT}}$ and the band bottoms of ${\delta}$, ${\delta_{CT}}$ show negligible $k_z$ dependence. These may be exploited to study the temperature dependence of the reconstruction in data taken with the same photon energy.

\subsection{Polarization dependence}

To further understand the electronic structure of Ca$_{1-x}$Pr$_{x}$Fe$_2$As$_2$, polarization dependent measurements were conducted at SLS to identify the orbital characters of band structure. Fig.~\ref{polar}(a) illustrates two types of experimental setup with linearly-polarized photons. The incident beam and the sample surface normal define a mirror plane. For the $p$ (or $s$) experimental geometry, the electric field direction ($\hat{\varepsilon}$) of the incident photons is parallel (or perpendicular) to the mirror plane. The matrix element of the photoemission process can be described by
\begin{center}
$\left| M_{f,i}^\textbf{\emph{k}} \right|^2 \propto \left| \left\langle \phi_f^\emph{\textbf{k}}\left| \hat{\varepsilon}  \cdot \emph{\textbf{r}} \right|\phi_i^\emph{\textbf{k}} \right\rangle \right|^2$
\end{center}
where $\phi_i^\emph{\textbf{k}}$ and $\phi_f^\emph{\textbf{k}}$ are the initial- and final-state wave functions. In our experimental setup, the momentum of the final-state photoelectron is in the mirror plane and $\phi_f^\emph{\textbf{k}}$ can be approximated by a plane wave. Therefore, $\phi_f^\emph{\textbf{k}}$ is always even with respect to the mirror plane. Thus considering the spacial symmetry of the Fe 3$d$ orbitals [Fig.~\ref{polar}(b)], when the analyzer slit is along the high-symmetry directions, the photoemission intensity of specific even (or odd) component of a band is only detectable with the $p$ (or $s$) polarized photons { \cite{Polar1, Polar2}}. For example, with respect to the mirror plane, the even orbitals ($d_{xz}$, $d_{z^2}$, and $d_{x^2-y^2}$) and the odd orbitals ($d_{xy}$ and $d_{yz}$) could be only observed in the $p$ and $s$ geometries respectively.

In the HT phase [Figs.~\ref{polar}(d1) and (e1)], $\alpha$ ($\beta$ and $\delta$/$\eta$) can be distinguished in the $p$ ($s$) geometry, indicating that their orbital characters should be even (odd). According to previous knowledge on iron pnictides {\cite{Polar1,Polar2}}, we can ascribe $\alpha$ to the $d_{xz}$ orbital, $\beta$ to the $d_{yz}$ orbital, while $\delta$/$\eta$ to the $d_{xy}$ orbital along these momentum cuts. The intensity of $\zeta$ band is weak in both $p$ and $s$ geometry, and thus it is hard to determine its orbital character. In the CT phase [Figs.~\ref{polar}(d2) and (e2)], $\alpha_{CT}$, $\beta_{CT}$, $\zeta_{CT}$ and $\kappa_{CT}$ show up in both geometries, suggesting that they are mixtures of both odd and even orbitals. The $\sigma_{CT}$ band can be resolved in the $p$ geometry, so they should be made of even orbitals. While $\delta_{CT}$ and $\eta_{CT}$ are present in the $s$ geometry, indicating that their orbital characters should be odd.

The orbital characters of the band structure is summarized in Fig.~\ref{polar}(f). It is reasonable to deduce that $\delta_{CT}$ and $\delta$/$\eta$ may have the same origin for their similar polarization and $k_z$ dependencies [see Fig.~\ref{KZH} and Fig.~\ref{KZL}]. On the other hand, $\alpha_{CT}$ and $\beta_{CT}$ probably originate from a strong mixing of $\alpha$ and $\beta$ during the remarkable electronic structure reconstruction, so that they appear in both $s$ and $p$ geometries.

\subsection{Temperature dependence}

\begin{figure}[t!]
\includegraphics[width=8.6cm]{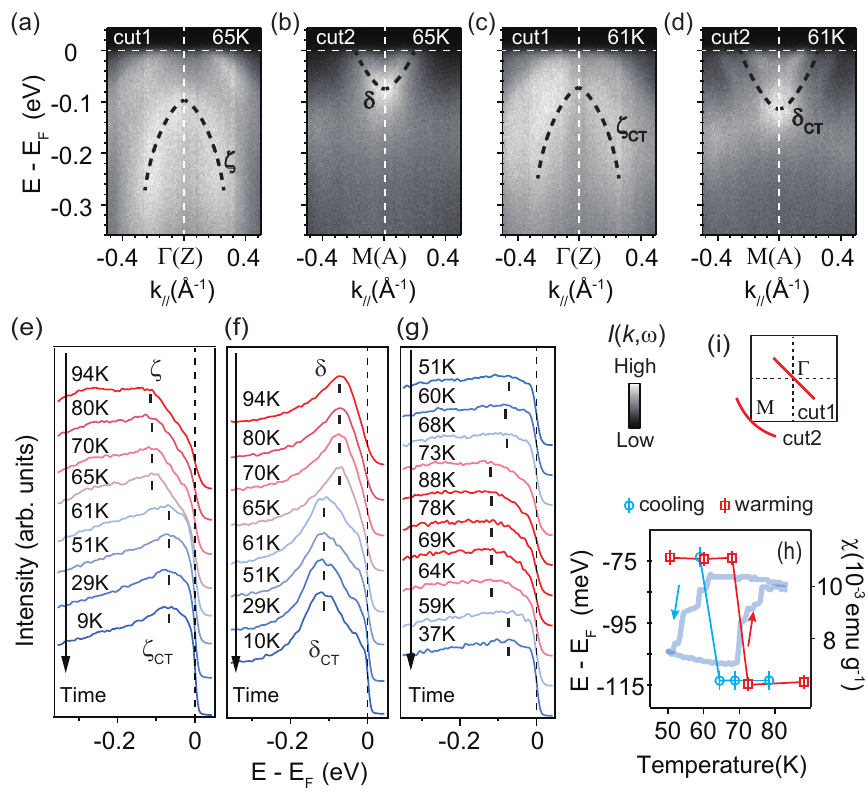}
\caption{(Color online) Temperature dependence of photoemission data. (a), (b) Photoemission intensity plots of Ca$_{1-x}$Pr$_{x}$Fe$_2$As$_2$ ($x$ = 0.15) measured along cut 1 and 2 respectively in the HT phase. (c), (d) The same for those the in CT phase as in panel (a) and (b). Dashed lines are guide of eye for band dispersions. (e), (f) The EDCs at $\varGamma$($Z$) and $M$($A$) for a cooling procedure respectively. Peak positions are indicated by short bars. (g) The EDCs of Ca$_{1-x}$Pr$_{x}$Fe$_2$As$_2$ ($x$ = 0.1) at the zone center [$\varGamma$($Z$)] for a warming-cooling cycle. (h) The summary of band tops of ${\zeta}$ [${\zeta_{CT}}$] obtained in panel (g), which exhibits a hysteresis. The DC susceptibility in Fig.\ref{lattice}(d) is overlaid for comparison (translucent blue lines). (i) Indication of cut directions in the projected two-dimensional Brillouin zone. All data were taken with randomly-polarized 21.2~eV photons.}
\label{Tdep}
\end{figure}

The detailed temperature dependence of photoemission data are shown in Fig.~\ref{Tdep}. The data collected on Ca$_{1-x}$Pr$_{x}$Fe$_2$As$_2$ ($x$ = 0.15) in a sample cooling procedure are presented in Figs.~\ref{Tdep}(a)-(f). The representative photoemission intensity plots along cut 1 and 2 [illustrated in Fig.~\ref{Tdep}(i)] are shown in Figs.~\ref{Tdep}(a) and (b) for the HT phase, while those for the CT phase are presented in Figs.~\ref{Tdep}(c) and (d). The corresponding EDCs at $\varGamma$($Z$) and $M$($A$) points are stacked in Figs.~\ref{Tdep}(e) and \ref{Tdep}(f) respectively. The peak positions of $\zeta$ ($\zeta_{CT}$) and $\delta$ ($\delta_{CT}$) change discontinuously across the CT transition, indicating that the reconstruction occurs abruptly.

Since the magnetic susceptibility shows a hysteresis loop, it is intriguing to investigate whether a similar hysteresis could be observed for the electronic structure. The EDCs at the zone center taken in a warming-cooling cycle of a Ca$_{1-x}$Pr$_{x}$Fe$_2$As$_2$ ($x$ = 0.1) sample are stacked in Fig.~\ref{Tdep}(g). The temperature dependence of band top of ${\zeta}$ (${\zeta_{CT}}$) band is summarized in Fig.~\ref{Tdep}(h). The clear hysteresis in the electronic structure reconstruction is qualitatively consistent with the bulk transport properties, which suggests the measured electronic structure somewhat reflects the bulk properties and the reconstruction is directly linked with the CT transition. Little deviation in the transition temperature is likely due to some sample variations.

\section{DISCUSSION AND CONCLUSION}

Previous theoretical works suggest that the remarkable change in the crystal structure across the CT transition, associated with the formation of As-As bonding  {\cite{non-magnetic_theory1, non-magnetic_theory2, As-As_bonding_1}}, quenches the magnetic moments of Fe and leads to the electronic structure reconstruction in the CT phase of Ca$_{1-x}$Pr$_x$Fe$_2$As$_2$.
However, compared with those of Ca(Fe$_{1-x}$Rh$_x$)$_2$As$_2$ and the annealed CaFe$_2$As$_2$, the reconstruction across CT transition in Ca$_{1-x}$Pr$_x$Fe$_2$As$_2$ is rather different. In Ca(Fe$_{1-x}$Rh$_x$)$_2$As$_2$ and the annealed CaFe$_2$As$_2$, the vanishment of hole cylindrical Fermi surface was argued to be intimately linked with the absence of superconductivity in the CT phase. However, the hole pocket still exists in the CT phase of Ca$_{1-x}$Pr$_x$Fe$_2$As$_2$, and the nesting condition still holds. This finding indicates that the absence of magnetic ordering and superconductivity in the CT phase is not necessarily correlated with the presence or absence of hole pocket.

Around the zone corner, four bands (${\kappa_{CT}}$, ${\eta_{CT}}$, ${\delta_{CT}}$ and ${\sigma_{CT}}$) are found in the CT phase of Ca$_{1-x}$Pr$_x$Fe$_2$As$_2$, while only one electron-like band is found in the CT phase of Ca(Fe$_{1-x}$Rh$_x$)$_2$As$_2$ and the annealed CaFe$_2$As$_2$. One possibility is that some of these bands might have a surface origin. However, the observed 1$\times$2 and 2$\times$1 surface reconstructions should contribute to the band folding at $X$ point rather than $M$ point {\cite{122_surface1,122_surface2}}, thus they are unlikely due to surface states. They are also unlikely caused by the weak magnetic ordering of praseodymium dopants at low temperatures {\cite{CaPr_NMR}}, since it was shown before that the ordering of Eu moment in EuFe$_2$As$_2$ does not cause noticeable change to its low energy electronic structure {\cite{122_ARPES2}}. Therefore, the unique band structure of Ca$_{1-x}$Pr$_x$Fe$_2$As$_2$ is likely intrinsic, but further band structure calculations are needed to investigate how these additional bands are caused by the CT structural transition.

Our data show that almost all the bands tend to be more dispersive in the CT phase than in the HT phase of Ca$_{1-x}$Pr$_x$Fe$_2$As$_2$. We applied parabolic-curve fittings to evaluate the effective mass of each band in these two phases in approximately the same $k_z$ plane. The effective mass of $\delta$ in the HT phase slightly reduces from 1.73 $\pm$ 0.02~$m_e$ to 1.6 $\pm$ 0.1~$m_e$ ($\delta_{CT}$) in the CT phase. More evidently, the absolute values of the effective mass of ${\alpha}$ and ${\beta}$ are 10.0 $\pm$ 0.4~$m_e$ and 4.1 $\pm$ 0.5~$m_e$  respectively. While in the CT phase, those values of ${\alpha_{CT}}$ and $\beta_{CT}$ are 5.7 $\pm$ 0.1~$m_e$ and 2.0 $\pm$ 0.1~$m_e$ respectively. The reduction of effective mass indicates the suppression of electronic correlations in the CT phase, which is also the case in CaFe$_2$As$_2$  {\cite{CaFe2As2_CT_ARPES_1, suppression_correlation_1, suppression_correlation_2, suppression_correlation_3}}. For iron-based superconductors, it was proposed in the strong-coupling pairing scenario that the superconducting pairing is mediated by the local antiferromagnetic exchange interaction {\cite{Strong-coupled}}, which is linked to the reasonably strong electronic correlations or narrow bandwidth {\cite{ZRY}}. Since the Fermi surface topology might be less relevant, it is most likely that the weakened correlations in the CT phase suppress the magnetic fluctuation, and subsequently push the system into the non-superconducting regime {\cite{ZRY}}.

In summary, we report detailed ARPES results on Ca$_{1-x}$Pr$_x$Fe$_2$As$_2$ ($x$ = 0.1 and 0.15) single crystals. Across the CT transition, the sizable change in the crystal structure leads to the drastic electronic structure reconstruction. Our results show discrepancies with the band calculations of CaFe$_2$As$_2$ and significant mixing of different orbitals in the CT phase, which still call for further theoretical investigations. Instead of the lack of Fermi surface nesting between electron and hole Fermi surfaces, we propose that the weakening of electronic correlations might be responsible for the absence of magnetic ordering and superconductivity in the CT phase.


\begin{center}
\textbf{ACKNOWLEDGMENTS}
\end{center}

Some of the preliminary data (not shown here) were taken at APE beamline of ELETTRA synchrotron light source. We gratefully thank Ivana Vobornik at ELETTRA, D. H. Lu and Y. Zhang at SSRL and Ming Shi at SLS for technical support. This work is supported in part by the National Science Foundation of China and National Basic Research Program of China (973 Program) under the grant Nos. 2012CB921402, 2012CB927401, 2011CB921802, 2011CBA00112, 2011CBA00106. D. W. Shen are also supported by the ``Strategic Priority Research Program (B)" of the Chinese Academy of Sciences (Grant No. XDB04040300). SSRL is operated by the U. S. DOE Office of Basic Energy Science.


\begin{references}

\bibitem{pnictide_SDW1} C. de la Cruz, Q. Huang, J. W. Lynn, J. Li, W. Ratcliff II, J. L. Zarestky, H. A. Mook, G. F. Chen, J. L. Luo, N. L. Wang, and P. Dai, Nature (London) \textbf{453}, 899 (2008)
\bibitem{pnictide_SDW2} M. Rotter, M. Tegel, D. Johrendt, I. Schellenberg, W. Hermes, and R. P\"{o}ttgen, Phys. Rev. B \textbf{78}, 020503(R) (2008)

\bibitem{induce_SC1} X. H. Chen, T. Wu, G. Wu, R. H. Liu, H. Chen, and D. F. Fang, Nature (London) \textbf{453}, 761 (2008)
\bibitem{induce_SC2} M. Rotter, M. Tegel, and D. Johrendt, Phys. Rev. Lett. \textbf{101}, 107006 (2008)
\bibitem{induce_SC3} K. Sasmal, B. Lv, B. Lorenz, A. M. Guloy, F. Chen, Y.-Y. Xue, and C.-W. Chu, Phys. Rev. Lett. \textbf{101}, 107007 (2008)

\bibitem{CaFe2As2_hydro1} A. Kreyssig, M. A. Green, Y. Lee, G. D. Samolyuk, P. Zajdel, J. W. Lynn, S. L. Bud'ko, M. S. Torikachvili, N. Ni, S. Nandi, J. B. Le\~{a}o, S. J. Poulton, D. N. Argyriou, B. N. Harmon, R. J. McQueeney, P. C. Canfield, and A. I. Goldman, Phys. Rev. B \textbf{78}, 184517 (2008).
\bibitem{CaFe2As2_hydro2} M. S. Torikachvili, S. L. Bud'ko, N. Ni, and P. C. Canfield, Phys. Rev. Lett. \textbf{101}, 057006 (2008).
\bibitem{CaFe2As2_hydro3} D. K. Pratt, Y. Zhao, S. A. J. Kimber, A. Hiess, D. N. Argyriou, C. Broholm, A. Kreyssig, S. Nandi, S. L. Bud'ko, N. Ni, P. C. Canfield, R. J. McQueeney, and A. I. Goldman, Phys. Rev. B \textbf{79}, 060510 (2009).

\bibitem{non-magnetic_theory1} T. Yildirim, Phys. Rev. Lett. \textbf{102}, 037003 (2009)
\bibitem{non-magnetic_theory2} W. Ji, X.-W. Yan, and Z.-Y. Lu, Phys. Rev. B \textbf{83}, 132504 (2011)
\bibitem{non-magnetic_experiment1} J. H. Soh, G. S. Tucker, D. K. Pratt, D. L. Abernathy, M. B. Stone, S. Ran, S. L. Bud'ko, P. C. Canfield, A. Kreyssig, R. J. McQueeney, and A. I. Goldman, Phys. Rev. Lett. \textbf{111}, 227002 (2013)
\bibitem{non-magnetic_experiment2} H. Gretarsson, S. R. Saha, T. Drye, J. Paglione, J. Kim, D. Casa, T. Gog, W. Wu, S. R. Julian and Y.-J. Kim, Phys. Rev. Lett. \textbf{110}, 047003 (2013)

\bibitem{CaFe2P2_QO} A. I. Coldea, C. M. J. Andrew, J. G. Analytis, R. D. McDonald, A. F. Bangura, J.-H. Chu, I. R. Fisher, and A. Carrington, Phys. Rev. Lett. \textbf{103}, 026404 (2009)

\bibitem{CaFe2As2_thermal_treatment} S. Ran, S. L. Bud'ko, D. K. Pratt, A. Kreyssig, M. G. Kim, M. J. Kramer, D. H. Ryan, W. N. Rowan-Weetaluktuk, Y. Furukawa, B. Roy, A. I. Goldman, and P. C. Canfield, Phys. Rev. B \textbf{83}, 144517 (2011).
\bibitem{CaFe2As2_doping} S. R. Saha, N. P. Butch, T. Drye, J. Magill, S. Ziemak, K. Kirshenbaum, P. Y. Zavalij, J. W. Lynn, and J. Paglione, Phys. Rev. B \textbf{85}, 024525 (2012).

\bibitem{Ca(FeRh)2As2_CT_ARPES} K. Tsubota, T. Wakita, H. Nagao, C. Hiramatsu, T. Ishiga, M. Sunagawa, K. Ono, H. Kumigashira, M. Danura, K. Kudo, M. Nohara, Y. Muraoka, and T. Yokoya, J. Phys. Soc. Jpn. \textbf{82}, 073705 (2013).
\bibitem{CaFe2As2_CT_ARPES_1} R. S. Dhaka, R. Jiang, S. Ran, S. L. Bud'ko, P. C. Canfield, B. N. Harmon, A. Kaminski, M. Tomi\'c, R. Valent\'i, and Y. Lee, Phys. Rev. B \textbf{89}, 020511(R) (2014).
\bibitem{CaFe2As2_CT_ARPES_2} K. Gofryk, B. Saparov, T. Durakiewicz, A. Chikina, S. Danzenb\"{a}cher, D. V. Vyalikh, M. J. Graf, and A. S. Sefat, Phys. Rev. Lett \textbf{112}, 186401 (2014)

\bibitem{CaFe2As2_no_hole_1} Y.-Z. Zhang, H. C. Kandpal, I. Opahle, H. O. Jeschke, and Roser Valent\'i, Phys. Rev. B \textbf{80}, 094530 (2009)
\bibitem{CaFe2As2_no_hole_2} M. Tomi\'c, R. Valent\'i, and H. O. Jeschke, Phys. Rev. B \textbf{85}, 094105 (2012)

\bibitem{CaPr_2SC_phase} B. Lv, L. Deng, M. Gooch, F. Wei, Y. Sun, J. K. Meen, Y.-Y. Xue, B. Lorentz, and C.-W. Chu, Proc. Natl. Acad. Sci. U.S.A. \textbf{108}, 15705 (2011)
\bibitem{CaPr_anisotropy} Y. Qi, Z. Gao, L. Wang, D. Wang, X. Zhang, C. Yao, C. L. Wang, C. D. Wang, and Y. Ma, Supercond. Sci. Technol. \textbf{25}, 045007 (2012)
\bibitem{CaPr_NMR} L. Ma, G.-F. Gi, J. Dai, S. R. Saha, T. Drye, J. Paglione, and W.-Q. Yu, Chin. Phys. B \textbf{22}, 057401 (2013)
\bibitem{CaPr_Hoffman} I. Zeljkovic, D. Huang, C.-L. Song, B. Lv, C.-W. Chu, and J. E. Hoffman, Phys. Rev. B \textbf{87}, 201108(R) (2013)
\bibitem{CaPr_local_inhomogeneity} K. Gofryk, M. Pan, C. Cantoni, B. Saparov, J. E. Mitchell, and A. S. Sefat, Phys. Rev. Lett. \textbf{112}, 047005 (2014)

\bibitem{CaFe2As2_growth} S. R. Saha, N. P. Butch, K. Kirshenbaum, J. Paglione, and P. Y. Zavalij, Phys. Rev. Lett. \textbf{103}, 037005 (2009)

\bibitem{ZRY} Z. R. Ye, Y. Zhang, F. Chen, M. Xu, J. Jiang, X. H. Niu, C. H. P. Wen, L. Y. Xing, X. C. Wang, C. Q. Jin, B. P. Xie, and D. L. Feng, Phys. Rev. X \textbf{4}, 031041 (2014)

\bibitem{122_surface1} E. van Heumen, J. Vuorinen, K. Koepernik, F. Massee, Y. Huang, M. Shi, J. Klei, J. Goedkoop, M. Lindroos, J. Van den Brink, and M. S. Golden, Phys. Rev. Lett. \textbf{106}, 027002 (2011)
\bibitem{122_surface2} Y.-B. Huang, R. Pierre, J.-H. Wang, X.-P. Wang, X. Shi, N. Xu, Z. Wu, A. Li, J.-X. Yin, T. Qian, B. Lv, C.-W. Chu, S.-H. Pan, M. Shi, and H. Ding, Chin. Phys. Lett. \textbf{30}, 017402 (2013)

\bibitem{CaFe2As2_ARPES} C. Liu, T. Kondo, N. Ni, A. D. Palczewski, A. Bostwick, G. D. Samolyuk, R. Khasanov, M. Shi, E. Rotenberg, S. L. Bud'ko, P. C. Canfield, and A. Kaminski, Phys. Rev. Lett. \textbf{102}, 167004 (2009)
\bibitem{Chenfei_ARPES} F. Chen, Y. Zhang, J. Wei, B. Zhou, L. X. Yang, F. Wu, G. Wu, X. H. Chen, and D. L. Feng, J. Phys. Chem. Sol. \textbf{72}, 469-473 (2011).

\bibitem{Polar1} Y. Zhang, F. Chen, C. He, B. Zhou, B. P. Xie, C. Fang, W. F. Tsai, X. H. Chen, H. Hayashi, J. Jiang, H. Iwasawa, K. Shimada, H. Namatame, M. Taniguchi, J. P. Hu, and D. L. Feng, Phys. Rev. B \textbf{83}, 054510 (2011)
\bibitem{Polar2} Y. Zhang, C. He, Z. R. Ye, J. Jiang, F. Chen, M. Xu, Q. Q. Ge, B. P. Xie, J. Wei, M. Aeschlimann, X. Y. Cui, M. Shi, J. P. Hu, and D. L. Feng, Phys. Rev. B \textbf{85}, 085121 (2012)

\bibitem{As-As_bonding_1} N. Colonna, G. Profeta, A. Continenza, and S. Massidda, Phys. Rev. B \textbf{83}, 094529 (2011)

\bibitem{122_ARPES2} B. Zhou, Y. Zhang, L. X. Yang, M. Xu, C. He, F. Chen, J. F. Zhao, H. W. Ou, J. Wei, B. P. Xie, T. Wu, G. Wu, M. Arita, K. Shimada, H. Namatame, M. Taniguchi, X. H. Chen, and D. L. Feng, Phys. Rev. B \textbf{81}, 155124 (2010).

\bibitem{suppression_correlation_1} Y. Furukawa, B. Roy, S. Ran, S. L. Bud'ko, and P. C. Canfield, Phys. Rev. B \textbf{89}, 121109(R) (2014)
\bibitem{suppression_correlation_2} J. Diehl, S. Backes, D. Guterding, H. O. Jeschke, and Roser Valent\'i, Phys. Rev. B \textbf{90}, 085110 (2014)
\bibitem{suppression_correlation_3} S. Mandal, R. E. Cohen, and K. Haule, Phys. Rev. B, \textbf{90}, 060501(R) (2014)

\bibitem{Strong-coupled} J. Hu and H. Ding, Sci. Rep. \textbf{2}, 381 (2012).

\end{references}
\end{document}